\documentstyle[11pt,aaspp4]{article}

\begin{document}

\title{A comparison of the optical properties of radio-loud and
radio-quiet quasars}

\author{P. Goldschmidt}
\affil{Imperial College of Science Technology and Medicine, 
Blackett Laboratory, Prince Consort Road,  London, SW7 2BZ, UK, p.goldschmidt@ic.ac.uk}

\author{M. J. Kukula}
\affil{Space Telescope Science Institute, 3700 San Martin Drive, Baltimore, MD
21218, USA and Institute for Astronomy, University of Edinburgh,
Royal Observatory, Blackford Hill, Edinburgh EH9 3HJ, UK, kukula@stsci.edu}

\author{L. Miller}
\affil{Nuclear and Astrophysics Laboratory, 
University of Oxford, Keble Road, Oxford, OX1 3RH, UK, l.miller1@physics.oxford.ac.uk}

\and

\author{J. S. Dunlop}
\affil{Institute for Astronomy, University of Edinburgh,
Royal Observatory, Blackford Hill, Edinburgh, EH9 3HJ, UK, jsd@roe.ac.uk}

\begin{abstract}

We have made radio observations of $87$ optically selected quasars at
5 GHz with the VLA in order to measure the radio power for these
objects and hence determine how the fraction of radio-loud quasars
varies with redshift and optical luminosity. The sample has been
selected from the recently completed Edinburgh Quasar Survey and
covers a redshift range of $0.3 \le z \le 1.5$ and an optical absolute
magnitude range of $-26.5 \le M_{B} \le -23.5$ ($h$,$q_{0}$
=$1/2$). We have also matched up other existing surveys with the FIRST
and NVSS radio catalogues and combined these data so that the optical
luminosity-redshift plane is now far better sampled than previously.
We have fitted a model to the probability of a quasar being radio-loud
as a function of absolute magnitude and redshift and from this model
infer the radio-loud and radio-quiet optical luminosity functions.
The radio-loud optical luminosity function is featureless and flatter
than the radio-quiet one. It evolves at a marginally slower rate if
quasars evolve by density evolution, but the difference in the rate of
evolutions of the two different classes is much less than was
previously thought.  We show, using Monte-Carlo simulations, that the
{\em observed} difference in the shape of the optical luminosity
functions can be partly accounted for by Doppler boosting of the
optical continuum of the radio-loud quasars and explain how this can
be tested in the future.

\end{abstract}


\section{Introduction}

Quasars were first discovered at radio frequencies
(\cite{sch63,hms63}) but subsequently only a small percentage were
shown to emit a significant fraction of their bolometric power in the
radio (\cite{san65}).  The distribution of radio luminosities appears
to be bimodal as shown by \cite{mpm90}, (hereafter MPM) Stocke et
al. (1992) and Hooper et al. (hereafter HIFH, 1995).  It is this
observation which has prompted many workers to classify quasars as
being `radio-loud' or `radio-quiet'.  
The usual division between the two classes is taken to be $\log10(P_{5
GHz}\rm{(W/Hz/ster)})=24$  which is the value that we have adopted in this
paper (we also assume $q_{0}=0.5$, $h=0.5$, $\Lambda=0$ throughout
unless specifically stated otherwise).  The results in this paper are
not too dependent upon the exact value of this dividing line although
we discuss below how they are affected by different definitions of
radio-loudness.

However, not all workers use
radio luminosity to define radio-loudness; Kellermann et al. (1989)
used $R$, the ratio of radio to optical luminosity.  This is a
physically meaningful parameter only if radio and optical luminosities are
{\em linearly} correlated. Peacock et al. (1986) and MPM argue that
this cannot be so, otherwise the fraction of optically faint
radio-louds should be much higher than measured.  No linear
correlation has been observed, at least for radio-loud quasars (Stocke
et al., 1993) although this might be dependent on spectral index as
Serjeant et al. (1997) claim a non-linear correlation in their large
sample of steep-spectrum quasars.

The debate over how to define radio-loudness is indicative of the fact
that we do not understand the link between radio and optical emission
of quasars.  It is still not known what the likelihood of a quasar
being radio-loud is as a function of optical luminosity and redshift.
The Palomar-Green Bright Quasar Survey (BQS, \cite{sch83}) is the most
studied sample of optically selected quasars at radio wavelengths and
the only one in which the sensitivity allowed the detection of some
radio-quiet quasars (\cite{kel89}) but even so, these data cannot
distinguish between the two opposing hypotheses in which the radio output
is either independent of optical luminosity or is correlated with it.
 
MPM used the BQS radio data together with VLA observations of higher
redshift quasars to show that the fraction of radio-louds increased as
redshift decreased.  More recently La Franca et al. (1994), using a
compilation of several radio studies, also showed that the likelihood
of a quasar being radio-loud increased as redshift decreased, but only
if the BQS data were included. Without these data, there is little
evidence for evolution in the fraction of radio-louds.  Both MPM and
La Franca et al. (1994) argue that the change in fraction of
radio-louds with redshift implies that the radio-loud population
evolve more slowly than the population as a whole. 
This could be an indication that the evolution is linked to the environments
of these objects.
Ellingson, Yee \& Green (1991)
have shown that radio-louds reside in much richer environments than
radio-quiets. If the former evolve slower than the latter then perhaps this
is because of an increased likelihood of the RLQs' fuel supplies being
`replenished' in rich environments. However an alternative explanation
is that inverse Compton scattering occurrs in the lobes of RLQs at high
redshift, thus decreasing these objects' radio luminosities. If there
were a correlation between radio and optical output then the
optical evolution of RLQs would be measured to be lower than that of RQQs.

HIFH used VLA observations of one third of the Large
Bright Quasar Survey (LBQS, \cite{hfc95}) and found that the number of
radio-loud quasars does not increase as redshift decreases, in direct
contrast to results using the BQS sample.  However, by using only the
LBQS, with the advantage of avoiding the usual mixture of different
surveys with different (usually undetermined) selection effects, this
paper suffered from the corresponding disadvantage of only spanning
about 1 magnitude in luminosity at each redshift.

It has been shown that the BQS quasar survey is incomplete
(\cite{gml92}) by a factor of 3.  Miller et al. (1993) have suggested
that this incompleteness is {\em not} random with respect to radio
properties, implying that the measured fraction of radio-louds in this
sample is biassed and cannot be used to determine the radio-loud
optical luminosity function. 

In this paper we use new radio observations of the optically-selected
Edinburgh Quasar Survey, and analyse these along with existing data to
quantify the relative numbers of radio-loud and radio-quiet quasars
(RLQs and RQQs) in optically-selected samples as a function of optical
luminosity and redshift. Our new data nearly doubles the number of
known RLQs derived from optical quasar surveys. Figs. 1 and 2 show the
coverage of the optical luminosity-redshift plane before and after the
addition of the radio data described in this paper, while Fig. 3 shows
the distribution of RLQs and RQQs in these samples. It can be seen
that the new data not only increase the number of known radio-loud
quasars selected from optical surveys but also samples far more of the
optical luminosity-redshift plane, allowing us to disentangle the
dependency of radio properties on both optical luminosity and
redshift.  The new data in Figs.~2 and 3 comprise 50 quasars from the
Edinburgh Survey (Miller et al. 1998, in preparation), 76 from the AAT
Survey (\cite{bfs90}) and 222 from the LBQS (\cite{hfc95}).

This paper is set out as follows. Section 2 describes the quasar
sample which was observed on the VLA, and the resulting data.  Section
3 describes how part of the AAT quasar survey (\cite{bfs90}) and LBQS
were matched up with the NVSS and FIRST 1.4 GHz surveys respectively,
and presents the resulting datasets.  Section 4 discusses and
quantifies evidence for bimodality in the radio luminosity
distributions.  Section 5 quantifies the the likelihood of a quasar
being radio-loud as a function of optical luminosity and redshift
using maximum likelihood.  It also presents results from a Monte-Carlo
simulation of the properties of radio-louds.  Section 6 discusses the
implications of this model fitting and Section 7 summarizes the main
conclusions of our work.

\section{The current sample}

A sample of 87 quasars were chosen at random from the recently
completed Edinburgh Survey (Goldschmidt 1993, Miller et al. 1998, in
prep.).  The redshift distribution spans $0.3 \le z \le 1.5$.  The
objects chosen all have optical magnitudes less than $B_{J} = 18.2$.

\subsection{Observations and data reduction}
 
6-cm continuum observations were made using the Very Large Array (VLA)
in the hybrid DnC configuration in 1993 December and 1994 January. A
dual-IF observing mode was used with two 50-MHz bands centred on
4.8226 and 4.8726~GHz.  Frequent observations of nearby point sources
were used to phase calibrate the target sources. Each quasar was
observed at two different hour angles in order to maximise coverage of
the {\it uv}-plane, with a total on-source integration time of
approximately 4 minutes per object.
 
We followed the standard calibration procedure within the AIPS
package.  Flux densities were calibrated relative to 3C286
($1328+307$), using the revised coefficients from the 1990 VLA
Calibrator Manual and leading to uncertainties in the derived fluxes
of around 5\%.  The two IFs were combined to give an effective
bandwidth of 100~MHz centred on 4.8476~GHz. The Fourier Transform was
carried out using a natural weighting scheme to maximise the
sensitivity in the final maps, giving a 3-sigma noise of roughly
0.2~mJy~beam$^{-1}$.
 
The synthesized beam of DnC-array at 6~cm is $\sim14''$ (FWHM) and the
useful field of view (FOV) is $9'$ (the primary beam of the
individual array elements causes severe attenuation outside this
region). Because of this large effective FOV many field sources
appeared in the final maps. We rejected as field sources any
detections which were more than $14''$ from the optical position of
the quasar (at the lowest redshift of the sample this would correspond
to a projected linear distance of 80~kpc), leaving us with 16 radio
detections and 71 upper limits.
 
The shortest baseline in this configuration is 35~m so the array will
only become `blind' to featureless extended emission on scales larger
than $5'$ which is much larger than any structures likely to be
associated with quasars in our redshift range.

\subsection{Results}
 
The results of the observations are detailed in Table 1. For objects
with a radio detection we give the measured flux and position of the
radio source, obtained by fitting a two-dimensional Gaussian function
with the AIPS task JMFIT.  We estimate that the radio positions are
accurate to $\sim0.1''$. For the brighter sources the uncertainty in
the flux measurement is dominated by the $\sim5\%$ calibration
uncertainty; for weaker objects the 3-sigma noise of
$\sim$0.2~mJy~beam$^{-1}$ dominates. Eight quasars in the Edinburgh
sample were already known to be Parkes, Molonglo, Green Bank or Texas
radio sources (these are indicated in the table) and for these objects
the fluxes in this paper are in very good agreement with the previous
measurements for those quasars measured at the same frequency.

Also, 37 of our objects are in common with the LBQS and have been
observed by HIFH, we have also included their determinations of the
8.4 GHz fluxes in the table, and spectral indices where relevant. The
mean spectral index for the few objects detected by both HIFH, and us
is 1, and the variance is large.

For non-detections we quote the optical position of the quasar and an
upper limit on the radio flux based on the 3-sigma noise measured over
the central part of the map.
 
Even with the rather coarse angular resolution of the
DnC-configuration, three quasars, 1303+020 1407-023, and 1430-004,
showed evidence for extended radio structure in the form of double or
triple sources straddling the optical nucleus (the high degree of
symmetry exhibited in each case makes it extremely unlikely that these
structures are merely due to a fortuitous alignment of background
sources).  For 1430-004 we took the flux of the central component as
the flux of the object.
 
%\subsubsection{Discussion of individual weirdo sources, esp. ones in Parkes}
%field sources? check map/check Parkes to find ID of powerful source.
%B double source, optical posn: 13 03 21.0 $+$02 05 32.0   check ID in Parkes
%7.738 mJy source at 14 29 12.966 -00 36 51.85
%B triple source optical position 14 30 10.0 -00 46 04.3 

\section{Matching existing optical quasar catalogues with recent
radio surveys}

There have been two major radio surveys carried out in the past few
years, the NRAO VLA Sky Survey (NVSS; \cite{con98}) and FIRST
(\cite{whi97}), that are capable of detecting all known radio-loud
quasars out to $z\sim 2$.  FIRST is in fact so sensitive (rms noise is
$\sim 0.15$ mJy) that it can detect many low-$z$ radio-quiet
quasars. In the next two subsections we describe matching up these
surveys with the AAT and the LBQS surveys respectively to increase the
statistics and also the coverage of the $L-z$ plane.

\subsection{Matching the AAT survey with NVSS data}

The AAT quasar survey (\cite{bfs90}) has an optical flux limit of $B
\le 21$, over 2 magnitudes fainter than the Edinburgh survey or LBQS.
No radio data are available for the AAT quasar survey.  We therefore
attempted to match up this survey with the NVSS to find radio
detections associated with the quasars.

The NVSS has a typical noise of $1\sigma=0.45$ mJy/beam and the
available catalogue is complete to a flux limit of $2.4$ mJy.  This
means that the survey can detect all known radio-loud quasars with
$z\le 1.5$ if they have luminosities $\log10(P) \ge 24$.
This is supported by Bischof \& Becker's recent (1997) matchup of the
NVSS catalogue with the V\'{e}ron-Cetty \& V\'{e}ron quasar catalogue
from which they conclude that the NVSS can detect most known
radio-loud quasars.

A radio detection in the NVSS was deemed to be an AAT quasar if it lay
within a $30$ arcsecond radius around the optical position.  There are
191 sources in the AAT sample within the NVSS area. Of these, 76 are
quasars with $0.3 < z < 1.5$ and $M_{B}\le -23$ so we consider this
subset to be a `complete' catalogue at both optical and radio
wavelengths. Using the criteria outlined above we matched up 5 of
these 76 quasars with NVSS detections.  Although $30$ arcseconds is
rather a large radius to use, the surface density of sources in the
NVSS ($\sim 60$ deg$^{-2}$) means that the probability of a random
NVSS source being identified as the radio counterpart to the optical
ID is $0.012$ and thus we expect that 1 out of our 76 quasars is
spuriously matched up. In fact, 4 out of the 5 matched quasars have an
offset between the optical and radio positions of $< 5$ arcseconds and
one has an offset of $29$ arcseconds.  It therefore seems likely that
this last source has been spuriously matched and we do not use its
associated radio flux in the next sections.  The coordinates, optical
and radio flux densities of the 4 reliably identified quasars are
given in Table 2.

Two of the above quasars are radio-loud, given our criterion.  This is
a very low fraction indeed ($3\%$), which, if genuine, has important
ramifications for the estimate of the radio-loud optical luminosity
function (OLF). We will discuss this in more detail in the following
sections.  Bischof \& Becker (1997) match up the first 2 of the
quasars in the above table (their catalogue only goes up to 21h30),
and do not match up any of the other AAT quasars. One of the
detections is a known radio-loud quasar, PKS 2203-18.

Is the low number of associations a consequence of errors in either or
both of the optical and radio positions? Condon et al. (1998) estimate
the error in the position of a faint NVSS source as being $\sim 5$
arcseconds.  The error in the optical positions is $\sim 1$ arcsecond
(\cite{bfs90}).  Therefore given the large search radius, errors in
the positions should not have affected the associations. Also, very
sensitive radio observations of the very faint survey ($b\le 22$,
\cite{zit92}) find 2 radio-loud quasars out of 55 (Gruppioni, private
communication) which confirms the low fraction of radio-louds in the
AAT survey.

We visually checked the associations by plotting NVSS fields for all
the AAT quasars within the NVSS area and overlaying the optical
positions. These visual searches confirmed the results above.

\subsection{Matching the LBQS with the FIRST radio survey}

In a similar manner we endeavoured to match the 1.4 GHz FIRST survey
(\cite{whi97}) with the LBQS. FIRST covers a smaller area to a deeper
flux limit than the NVSS; the 1$\sigma$ noise is on average $0.15$
mJy, and the available catalogue is complete to 1 mJy. Therefore we
estimate it should be capable of detecting radio-loud quasars out to
$z\ge 2.2$.  HIFH and Visnovsky et al. (1993) present VLA measurements
of $1/3$ of the quasars in the LBQS, however as HIFH show, the quasars
chosen for the VLA observations are more luminous and at higher
redshifts than the quasars not observed. Therefore by matching up the
LBQS with FIRST we should be able to extend coverage of the L-z plane
to slightly fainter luminosities, as well as improving the statistics.

The average surface density of the sources in the FIRST is
90/degree$^{2}$, and we chose a 10 arcsec radius around each LBQS
position within which to search for coincident radio sources. The
likelihood of a random radio source being matched up with the LBQS
quasar is $2.2\times 10^{-3}$. There are 222 LBQS quasars within the
FIRST survey area so we therefore expect 0.5 spurious matches.

We found a total of 28 detections out of the 222 quasars (Table 3).
Two of these are known to be  Parkes sources; PKS 0256-005 and PKS 0056-00.  
Radio
fluxes for 34 of these objects have already been reported in HIFH.  Of
these 34 objects, 4 are detected by both FIRST and HIFH, 1 object is
undetected by HIFH but FIRST detects it, and 2 objects are detected by
HIFH and not by FIRST.  These latter objects are radio-loud but at
$z=3.281$ and $z=3.015$ and therefore their radio luminosities are
just below the flux limit of FIRST (depending upon our assumed
spectral index), {\em i.e.} we expect FIRST to be incomplete at these high
redshifts.  The rest of the 27 quasars in common are undetected by
both surveys.  This comparison indicates that the low number of
radio-loud quasars that we find in the LBQS-FIRST match-up is genuine
and not due to incompleteness in FIRST.

%In order to test
%further the possibility of spurious matches we compared the observed offsets
%between the optical and radio positions, with the expected distribution for
%random associations. Fig. ?? (NOT DONE YET) shows the observed and expected distributions,
%it is clear that the observed distribution is much more strongly
%clustered around zero than the expected and so we are confident that our
%associations are genuine.

Many of our FIRST associations are in fact low redshift radio-quiet quasars
and the fraction of radio-louds is slightly lower than in HIFH who
estimate that $10\%$ (27 out of 256) of their quasars are radio-loud,
whereas we estimate that $7.7\%$ (17 out of 222) of ours are
radio-loud.  The difference is not statistically significant. However
it could be due to the fact that the LBQS quasars that we have matched
up are on average fainter and at higher redshifts than the sample in
Hooper et al, as shown in Fig. 2.

\subsection{Coverage of the optical luminosity-redshift plane}
 
Here we briefly summarize the main properties of the samples
used in this paper.

Fig. 2 shows the coverage of the optical luminosity - redshift plane
achieved to date in published radio surveys of optically-selected
quasars, together with the data discussed in Sections 2 and 3.1 and
3.2 above.  We do not include the BQS survey on this figure because of
the doubts mentioned above about the possible bias in this survey
towards radio-loud quasars; this is discussed in more detail in
Section 5.1.  The plot highlights the problem of the
luminosity-redshift correlation which is endemic in flux-limited
surveys, with the result that large regions of the plane remain
unsampled. This patchy coverage is of particular concern in studies
such as the current one, in which we wish to separate the effects of
redshift and absolute magnitude on quasar radio properties. In order
to do this with any degree of confidence we need to be able to compare
objects of similar absolute magnitude at different redshifts and vice
versa.  It can be seen that the MPM data breaks this
luminosity-redshift degeneracy at $z\sim 2$ and now, using the
combined data from AAT, LBQS and Edinburgh we can break the degeneracy
at lower $z$ as well.
 
The LBQS, although by far the largest sample to have been surveyed in
the radio is not ideally suited to this purpose when considered in
isolation; for a particular redshift the spread in absolute magnitude
amounts to only a magnitude or so (Figure 1) making it impossible to
compare objects of equivalent optical luminosity over redshift
intervals of $z \geq 0.2$. However this is a very large sample; 359
LBQS quasars were observed with the VLA by Hooper et al. (1994, 1995)
and Visnovsky et al. (1993) with $0.2 < z < 3.4$ and $-23 > M_{B} >
-28$.
 
The samples of MPM consist of 105 quasars and provides fairly uniform
coverage of the region $1.8 < z < 2.4$ and $-25 > M_{B} > -27.5$.  MPM
used the VLA in C-configuration at 5~GHz (6~cm; 5-arcsec resolution)
to observe 99 quasars from the quasar surveys of Osmer (1980) and
Osmer \& Smith (1980), and included a further 6 quasars from Smith \&
Wright (1980).  With typical $3\sigma$ sensitivities of $\sim
0.6$~mJy~beam$^{-1}$, this corresponds to a radio luminosity of
$10^{24}$W~Hz$^{-1}$ster$^{-1}$.

Schneider et al. (1992) present radio data for 22 quasars at high
redshift ($3.2 < z < 4.7$), covering a similar range of absolute
magnitudes ($-25.0 \geq M_{B} \geq -27.5$) to MPM. 

La Franca et al. (1994) observed 23 optically selected quasars from
the SA94 survey (\cite{flf92}) and detected 4 of them. They achieved a
$1\sigma$ noise of $\sim 0.05$ mJy and find their data are consistent
with a bimodal radio distribution out to moderately high redshifts.

It is extremely important to point out that many of the observations
made by other authors and used for comparison in this paper were not
carried out using the same wavelengths and magnitude systems.  For the
purposes of Figs.~2 and 3, and in all subsequent discussion, we have
transformed the original data to our preferred units: radio flux
density at a rest wavelength (after $k$-corrections) of 6~cm (5~GHz)
and absolute magnitudes measured in the $B$ Johnson system.  The
assumptions made during this process have important implications, and
merit some discussion.  Because Figs.~2 and 3 cover a large range of
redshifts and because not all of the quoted radio observations were
carried out at a wavelength of 6~cm (5~GHz), the assumed form of the
quasar spectrum is doubly important as it affects both the
transformation from one {\it observed} wavelength to another and the
transformation to the same {\it rest} wavelength in each case. For the
purposes of $k$-corrections and the transformation of radio flux
density between different observing bands we take the optical and
radio spectral slopes to be $\alpha_{O}$ and $\alpha_{R} = 0.5$, where
flux density $S_{\nu} \propto \nu ^{-\alpha}$.  The few objects for
which we possess spectral information have rather flat spectra with
$\alpha \le 0.5$ (see above) but the scatter is large and therefore
the data are reasonably consistent with this assumption and do not
justify the use of a different value.  However, we note that a
significantly steeper or flatter spectrum might conceivably affect the
fraction of objects classified as radio loud or radio quiet, or raise
the effective luminosity limit of a survey to a level at which it is
impossible to distinguish between radio-loud and radio-quiet using the
criterion described above.

For the optical luminosities we used $M_{B}$ where $B$ refers to the
Johnson system. This is not the natural system for most of the quasar
samples discussed above and hence we transformed the apparent $B_{J}$
magnitudes in the Edinburgh survey assuming the transformations in
Blair \& Gilmore (1982) and a mean $B_{J}-V$ colour of $0.18$ to give
$B = B_{J} + 0.06$. For the AAT survey we used the transformation
given in Boyle et al. (1988); $B = b -0.1$.  For the LBQS quasars we
used the absolute magnitudes given in HIFH which are transformed to
the $B$ system.  For the LBQS quasars in the FIRST survey we used the
same transformation as for the Edinburgh quasars.  The MPM quasars
already have apparent magnitudes quoted at the rest-frame wavelength
of $1475$\ \AA \ and thus we transform them to $B$ using
$B=m_{MPM}+0.23$ which is correct at the average redshift of the
sample.

There are clear inconsistencies involved in transforming all these
different surveys to a common system. Ideally one would like to use
the natural system without any transformation as this only increases
the uncertainty.  Also, an estimate of the {\it continuum} flux
without any contribution from emission lines would be ideal,
especially for this type of study, as it is possible that radio-louds
and -quiets might have different emission line strengths. In spite of
this caveat, the uncertainties involved in transforming from one
system to another are most likely much smaller than the random errors
in the measured fluxes which could be as high as $15\%$.

A comparison of Figs. 1, 2 and 3 show how the radio information on the AAT
survey covers a part of the $M_{B}$-z plane that has not been sampled
before. They also show how increasing the radio coverage of the LBQS
improves the statistics of brighter quasars.  Both these improvements
allow us to determine the OLF of radio-louds over a wider range of
redshifts.  We do this below, in Section 5.

\section{Evidence for bimodality in the radio luminosity distribution}

Strittmatter et al. (1980) and MPM argued that the distribution of
radio luminosities is bimodal. If true, this has important
implications for our understanding of the origins of radio emission in
quasars. In this section we look at how to test for bimodality and
what conclusions can be drawn from the results. Figs. 4 and 5 show the
optical and radio luminosity distributions for the data used in this
paper separated into two redshift bins. It can be seen that there is some
evidence for bimodality but this is not overwhelming, particularly at low
redshift where there is little evidence for a `gap' at $log(P) \sim 24$.

It is not trivial to quantify the likelihood of a dataset being
bimodal.  One maximum-likelihood-based method is the Lee statistic
(Lee, 1979, Fitchett, 1988).  This looks at the variances in subsets
of the data and compares the weighted sum of these variances with that
for the whole dataset. If there is a way of dividing the data such
that the variances in each subset are very much smaller than for the
dataset as a whole then this implies that the distribution is
bimodal. However to quantify the significance of this one needs to
assume a parametric unimodal distribution and estimate (from
Monte-Carlo simulations) the likelihood of the Lee statistic
describing this as bimodal. Clearly the major problem is that most of
the data consists of upper limits and hence we have little idea of
what sort of parametric distribution to assume.  A reasonable approach
seems to be to use the upper limits as detections and fit a gaussian
with a uniform tail to high radio luminosities. Stocke et al. (1992)
and HIFH used similar approaches.
 
The Lee statistic was calculated in 2 redshift bins; $0.5 \le z < 1.5$
and $1.5 \le z \le 2.5$. We then estimated the significance of the
measured Lee ratios from Monte-Carlo simulations of a unimodal
distribution consisting of a Gaussian with a uniform tail to high
luminosities. In neither of the 2 redshift bins were the measured Lee
ratios estimated to be significantly different from the distribution
of Lee ratios that we derived from the Monte-Carlo realisations. We
conclude that there is insufficient evidence to show that the radio
luminosity distribution is indeed bimodal. HIFH reach the same
conclusion and emphasize the need for more data. 

\section{Luminosity and redshift distributions of radio-loud quasars}

\subsection{Does radio-loud fraction vary with redshift?}

In this and the next section we attempt to separate out the effects of
$z$ and $M_{B}$ upon the fraction of radio-loud quasars; f(RL).  By
combining the current sample with the LBQS we obtain 124 quasars in
the absolute magnitude interval $-25.5 \ge M_{B} \ge -26.5$, covering
a redshift range of $\Delta z=0.6$ centred on $z = 0.9$. These objects
have the same range of optical luminosities as the high-redshift ($z
\simeq 2$) sample of MPM (assuming $\Omega=1$), allowing us to test
for signs of evolution in the fraction of radio-loud sources between
the two epochs. We did this by calculating the numbers of radio-loud
and -quiet quasars in each of the two redshift bins and carrying out a
contingency table analysis.  The fraction of RLs is $18\%$ at low $z$,
dropping to $8\%$ at higher $z$.  However this is not statistically
significant; $\chi^2=2.06 $ with 1 degree of freedom, and thus the
fraction of radio-loud quasars to total number of quasars in each bin
is consistent at the $15\%$ level.

We then included the VLA data on very high redshift quasars from
Schneider et al. (1992), and fitted a simple parametric model to the
fraction of radio-loud quasars, f(RL), as a function of redshift. We
chose to model the data as
\begin{equation}
	\frac{n(RL)}{n(RQ)+n(RL)} = A(1 + z)^{\gamma}
\end{equation}
where $\gamma=0$ if the fraction stays constant with redshift. The
best fit parameters are $\gamma=-2.16$ and $A=0.75$ (see Figure~6 for
more details), indicating a slight decrease in f(RL) with increasing
redshift. However a constant fraction of radio-loud quasars with
redshift cannot be ruled out and is consistent with the data at the
$16\%$ level.

We have also plotted the estimated fraction of radio-louds at low
redshift from the BQS sample within the same magnitude range on Figure
6.  It can be seen that the BQS sample has a higher fraction of
radio-louds, which is different from the model at the $99.8\%$ level.

To summarize, although there are hints that f(RL) increases as $z$
decreases for all $M_{B}$ within the $z$ range $0.6 < z < 2.5$ this is
not statistically significant in the narrow $M_{B}$ range that we have
chosen.  A more direct approach is to try and determine the fraction
of radio-louds as a function of $M_{B}$ in different redshift slices
and we do this in the next section.

\subsection{Radio-loud fraction as function of optical luminosity}

Because we have managed to increase coverage of the $M_{B}-z$ plane by
using the NVSS and FIRST radio surveys, we can attempt to estimate the
fraction of radio-loud quasars as a function of absolute magnitude in
redshift slices without being too troubled by the usual
luminosity-redshift degeneracy.

We used a subset of the data from the Edinburgh, LBQS, AAT, SA94 and
Schneider et al.  samples in two redshift bins, and measured the
number of radio-loud quasars as a function of magnitude in each
bin. Fig.~7 shows this fraction for each bin together with the
best-fit model, discussed below.  The error bars show the 1$\sigma$
errors, which were calculated using $\sigma^2(f)=f(1-f)/N$ where $f$
is the RLQ fraction and $N$ is the number of quasars per bin.

Below we provide a more detailed analysis of these plots but here we
point out the main features.\\(a) In each redshift bin the fraction of
radio-louds increases with luminosity.  This implies that the
radio-loud OLF is slightly flatter than the total OLF for most
luminosities.\\(b) There is considerable overlap in the luminosities
sampled, and thus we can say that the fraction of radio-louds is
marginally higher within the luminosity range $-26.5 < M_{B} < -24.5$
at lower redshifts.
%although this does not appear to be statistically significant.

Together, these two observations are consistent with a model in which
the radio-loud OLF is flatter than the total OLF but evolves at the
same rate, such that in a given $M_{B}$ slice as $z$ decreases the
fraction of radio-louds increases, as marginally indicated by the
data. However, the data also imply that the RL OLF cannot be much
flatter than the total OLF because the rate of change with $z$ is only
marginal.  Below we compare such a model to the data and discuss its
implications.

\subsection{Model fit to radio-loud optical luminosity function}

We can use the data to fit models to the optical luminosity function
of radio-louds (and -quiets). We chose to do this using maximum
likelihood (ML).  The main advantage of this method over $\chi^{2}$ is
that it does not require the data to be binned first (see Marshall et
al. 1984 for a detailed description of using ML to estimate best-fit
parameters of luminosity functions).  We formulated the likelihood
function in the following way, similar to that in La Franca et al.
(1994), in which it is assumed that the total (RL+RQ) OLF is {\it a
priori} well-determined and therefore one can use the {\em relative}
numbers of RLs (as a function of optical luminosity and redshift) to
estimate the RL OLF. This is rather different in spirit from the
likelihood function of Marshall et al. (1984), which used the {\em
absolute} probability of detecting a quasar.  We cannot use that
method directly as we have, on the whole, only incomplete information
on the radio properties of each survey and hence cannot determine the
absolute normalisation of the RL OLF without taking into account the
total number of quasars that have been observed.  If $p$ is the
probability of a quasar being radio-loud then we can formulate the
likelihood function as
\begin{equation}
{\cal L} = \prod_{i=1}^{N_{RL}}p(M_{B},z)\prod_{i=1}^{N_{RQ}}(1-p(M_{B},z))
\end{equation}
and we choose to model $p$ as an arbitrary function of $M_{B}$ and
$z$;
\begin{eqnarray}
& & p(M_{B},z)dM_{B}dz = \nonumber \\
& &  (1 + z)^{C} 10^{A + B(M_{B}-M^{*}) + D(M_{B}-M^{*})^2}
dM_{B}dz.
\end{eqnarray}

We have to make a choice on the parameterization of the total OLF. As
has been discussed in recent papers (Hewett et al. 1993, Miller et
al. 1994, Koehler et al. 1997, Goldschmidt \& Miller, 1998) the Pure
Luminosity Evolution (PLE) model (Marshall et al. 1984, Boyle et al. 1988) -
in which the shape of the luminosity function remains invariant at all
redshifts and in which the evolution is parameterised by the
characteristic luminosity $M_{*}$ increasing with redshift - has been
shown to be a poor fit to the more recent bright quasar surveys.
Although it is not clear how to parameterize the `current' OLF, the
above authors have all shown using 3 independent surveys that the
shape of the OLF changes with redshift in that it is flatter at low
$z$. At $z< 1$ the OLF is a featureless power law.

Hewett et al. (1993) present a parameterization of the OLF for all
quasars (radio-loud and radio-quiet) which is based on the optically
selected LBQS.  This is a more arbitrary model than PLE and clearly
cannot be interpreted as having any physical meaning. We combined our
model fit to the fraction of RLQs with the Hewett et al.
parameterization of the OLF to give the OLFs of RLQs and RQQs
separately.

We achieved a satisfactory fit (a probability of $15\%$ from $\chi^2$)
to the data with the following parameters, $A=0.50 \pm 0.05, B=-0.56
\pm 0.02, C=-2.12 \pm 0.18, D=0.06 \pm 0.01, M^{*}=-28 \pm 0.10$; see
Fig. 7 for a comparison of the model and data in two redshift
bins. The errors on the parameters were estimated assuming a $\chi^2$
distribution for $\Delta(-2\log{\cal L})$.  From our estimate of
$p(M_{B},z)$ we can now derive the radio-loud luminosity function
(since it is defined as $\Phi_{RL} =p \Phi_{total}$). We discuss this
further in Section 6 below.

\subsection{Does Doppler boosting affect the optical luminosities of radio-loud
quasars?}

Above we showed that the best-fit model to the RLQ OLF is one which is
skewed to higher luminosities than that of RQQs. Here we discuss a
possible physical mechanism for the difference in the optical
luminosity distributions.

In the orientation-based unification scheme (OBUS) the observed
differences between radio-loud quasars and radio galaxies can be
explained in terms of the differences in viewing angle. According to
the OBUS objects that are viewed at large angles to the jet axis have
more obscured optical continua and broad line regions.  Objects viewed
at small angles to the jet axis offer an almost unobscured view of the
central region and inner regions. These latter objects are also more
likely to suffer from Doppler boosting of the emission in the jet.  It
is this Doppler-boosted jet emission that we are concerned with here.

We suggest that this Doppler boosting may also affect the {\em
optical} emission and hence the optical luminosity distribution of
RLQs. Given a jet speed $v=\beta c$ and corresponding $\gamma$,
together with a viewing angle $\theta$ from the jet direction and
intrinsic spectral index $\alpha$, the effect upon the observed flux
$S_{obs}$ given the intrinsic flux $S_{int}$ is;
\begin{equation}
M = 
\frac{S_{obs}}{S_{int}}=
\left[\frac{1}{\gamma}\frac{1}{(1-\beta\cos(\theta))}
\right]^{3+\alpha}.
\end{equation}

The effect upon the observed fraction of radio-louds is a convolution
of the above distortion with the intrinsic luminosity distribution of
radio-louds. We modelled this using a Monte-Carlo simulation, with the
following assumptions;\\ a) We assumed that roughly $10\%$ of the
observed continuum flux at optical wavelengths is due to synchrotron
emission.\\ b) The maximum boost factor allowable is 10, implying that
the flux of the object is doubled. There needs to be an upper limit
factor on the possible range of boost factors, otherwise the
simulation might produce an excess of BL Lac objects, compared with our
observations.\\ c) We assumed a uniform distribution for the jet
speed; $0.1 \le \beta \le 0.95$.\\ d) We used, as the intrinsic RLQ
OLF, the fit to the RQQ OLF. In the absense of beaming this implies
that the fraction of RLQs is constant with $M_{B}$.\\ e) The range of
viewing angles is $0 \le \theta \le 45$ degrees.

Fig. 8 shows the results of this simulation. The fraction of RLQs is
clearly a function of optical luminosity and we can understand this as
being due to a combination of faint objects being beamed and appearing
brighter, and due to the finite slope of the luminosity function.
Although the fraction of RLQs does increase with luminosity in this
simulation, the dependency of f(RL) upon $M_{B}$ is not so strong as
in our data and thus we conclude that Doppler beaming may be
responsible for part of the observed dependency of f(RL) upon $M_{B}$
but cannot account for all of it, as this would predict that all of
the most optically luminous objects are blazars.

\section{Discussion}

Figure 9 shows the derived optical luminosity functions for both RLQs
and RQQs over the luminosity ranges for which we have information on
the radio classes, given the estimated probability of a quasar being
radio-loud as derived in Section 6.  It can be seen that the RLQ
OLF appears to be flatter and hence skewed to higher luminosities than
that for RQQs. We have already discussed a possible physical cause of
this flattening in the previous section. The next obvious question to
ask is do both OLFs evolve at similar rates?  Unfortunately it is not
easy to see how to answer this question, principally because, as we
discussed above, the total OLF changes shape. Because of this
there is no obvious `feature' in the OLF that can be traced at
different redshifts and hence there is no straightforward
parameterization of the evolution. If we consider density evolution then
the RLQs evolve more slowly than the RQQs, as marginally indicated by the
data in Section 5.  Without more information on the lifetimes of the
objects, or more physically detailed models of evolution, one cannot
say anything more, apart from noting that the lack of evolution
between $1.5 < z < 2$ shows up in {\em both} classes of objects. This
is a clear indication that the {\em onset} of optical evolution is a
universal phenomenom and not directly related to radio power, whereas
the {\em subsequent} evolution may well be a function of radio power.

%This is interesting because of the correlation between radio luminosity and
%richness of environment as indicated by e.g. Ellingson et al. (1991).
%This might be an indication that {\em optical} evolution is not dependent
%upon environment, although clearly we need to be able to constrain the
%OLFs of both classes of quasar better before we can substantiate such claims.

The Doppler boosting mechanism discussed above may be tested using
both radio and optical data. Although, as argued previously, the
effect upon individual quasar SEDs is likely to be rather small,
overall we can compare the average SEDs of RLQs and RQQs provided that
they are well matched in {\em observed} optical luminosity. The
beaming hypothesis predicts that the emission lines of the RLQ SED
will have smaller equivalent widths than those of RQQs because of the
extra beamed optical continuum. Using the radio properties to test
this hypothesis will require not only detections of the RQQs but also
spectral indices, as the hypothesis predicts that the radio spectra of
RQQs will be steeper those of RLQs. Also, with the current data we
have not been able to sub-divide the RLQs on the basis of their
spectral indices, however, if we restricted ourselves to the
steep-spectrum objects, we would expect a more similar optical
luminosity distribution to the RQQs since beaming should not be
observed for these objects. Serjeant et al. (1997) are addressing this
with their sample of radio-selected steep spectrum quasars.

There are various other plausible causes for the difference in OLFs
for the two radio classes. One obvious difference between the two
classes is the difference in host galaxies. This subject has a
contentious history, in that for a long time it was an article of
faith in AGN studies that radio-loud AGN resided in elliptical hosts,
and radio-quiet AGN resided in spiral hosts. The evidence for this
comes from very low redshift studies of Seyferts and radio
galaxies. However until recently the evidence that this dichotomy
extended to higher optical luminosities and/or redshifts was totally
lacking. The recent papers by McLeod \& Rieke (1995) and Taylor et al
(1996) using NIR imaging of $z \le 0.4$ RQQs and RLQs shows that the
distinction between the hosts breaks down. In particular Taylor et al.
found that the fraction of RQQs in elliptical hosts increases with
optical luminosity.  McLeod \& Rieke showed that there appeared to be
a lower limit to the luminosities of the hosts which was linked to
that of the resident quasar.  How does this fit in with our studies of
OLFs? The answer is linked to Peacock et al.'s (1986) original reason for the
lack of optically faint RLQs. If RLQs at faint luminosities are more
likely than RQQs to be in large, luminous ellipticals then there are
{\em two} ways that they could be selected against in optical
surveys. The first is because of morphology, optical surveys usually
require that quasar candidates are not extended (because of the
photometric calibration on Schmidt plates), and thus an optically
faint quasar in a luminous host galaxy is more likely to be
discriminated against than one in a smaller and/or fainter host
galaxy. The second possibility is that early-type galaxies appear to
be redder than late-types and thus the overall colour of the
AGN-plus-host system may not be UVX enough to satisfy the selection
criteria of the survey.

However, the {\em shape} of the dependency of radio-loud fraction on
optical luminosity may not be fully consistent with the above picture,
in that one would expect the effect of incompleteness due to large red
hosts to kick in at faint luminosities, whereas the fraction of RLQs
remains roughly constant over a wide range of $M_{B}$ and increases
sharply only at the highest luminosities sampled.

The observed difference in the OLFs could also be due to an underlying
correlation between radio and optical luminosities combined with our
definition of radio-loud as being above a cut-off in radio
luminosity. The putative correlation would then mean that the observed
optical luminosities of the RLQs chosen in this way are skewed to high
values and hence the OLF is observed to be flatter than the true
distribution.

%We conclude that the model of Doppler beaming is more consistent with the data.
%A more complicated model involving dust obscuration for various types of AGN
%may well be required, as indicated by Baker \& Hunstead (1996), but currently,
%the data are not good enough to constrain such models.

A separate but related question is; are RLQs responsible for the
flattening of the total OLF seen at $z\le 1$? As discussed above,
recent studies of the OLF show that it flattens at low redshifts. This
could in principle be entirely due to RLQs having a relatively high
space density at low $z$, compared to the population as a whole. But
the change in power-law index of the OLF means that the space density
of $z\le 1$ quasars is roughly a factor 3 higher than indicated by the
uniform PLE model. For RLQs to be responsible for this they would have
to make up the bulk of the population (at least $60\%$) at low
redshifts and this is not what we find.

\section{Conclusions}

We have acquired radio data for optically selected quasars from 3
different surveys and thus extended coverage of the optical
luminosity-redshift plane for radio-loud and -quiet quasars. These new
data have weakened the degeneracy between luminosity and redshift and
allowed us to model the optical luminosity functions and their
evolutions for the two classes of quasar.  We note that this study
depends on the combination of several different quasar samples, all
with different primary selection criteria and amounts of bias. The
cumulative effect of these biases is almost impossible to quantify,
but we stress that until much larger, optically-selected quasar samples
become available this remains the only way of obtaining sufficient
coverage of the $M_{B}-z$ plane.  Our conclusions can be summarized as
follows:

\noindent
(1) Radio luminosity distribution: In agreement with previous work,
the distribution of radio luminosities gives the impression of being
bimodal, but a unimodal distribution cannot be ruled out with any
significance from the results of the maximum-likelihood based Lee
test.

\noindent
(2) Radio-loud fraction: The fraction of radio-louds decreases with
optical luminosity, but there is little evidence for strong evolution
of this fraction with redshift.

\noindent
(3) Optical luminosity functions: It appears that, at a given
redshift, the optical luminosity function for radio-loud quasars is
skewed towards higher luminosities and is flatter than that for
radio-quiets. The optical luminosity function for the latter shows a
change in shape, similar to that seen in the total optical luminosity
function, although more radio data is needed for the most optically
luminous quasars in order to fully substantiate this.  Monte-Carlo
simulations of radio-loud quasars with the {\em same} intrinsic
optical luminosity function as radio-quiets show that if $10\%$ of the
optical emission for RLQs is synchrotron emission from the jet, then
Doppler beaming can be responsible for the some of the {\em observed}
difference in optical luminosity functions for the two classes of
object.

\noindent
(4) Evolution of RLQs and RQQs:  The space density of radio-loud
quasars evolves more slowly than that of radio-quiets at a given
optical luminosity, although the evolution for both class of quasar
slows similarly at $z \sim 2$.  What this implies for our
understanding of the {\em physical} causes of the evolution remains
unclear.

\acknowledgements
The Very Large Array (VLA) of the National Radio Astronomy Observatory
is operated by Associated Universities, Inc., under a cooperative
agreement with the National Science Foundation.  PG and MJK acknowledge
PPARC support. Data reduction was partially carried out on
STARLINK. We thank Steve Serjeant for useful conversations.
We are grateful for the referee's comments which improved this paper.
%We are also grateful to Jim Condon and Richard White and
%co-workers for making available the FIRST and NVSS radio surveys to the
%community.

\clearpage

% now figures..
\clearpage
\figcaption[fig1.eps]{The optical luminosity-redshift plane, showing
the distribution of quasars in optically-selected samples for which
radio data was available prior to this paper, including the objects
from the Palomar-Green (PG) Bright Quasar Survey.
\label{fig1}}
 
\figcaption[fig2.eps]{Optically-selected quasar samples for which
radio data is now available, including the new data described in this
paper. For reasons outlined in the text, we have not included the BQS
quasars. Note how the new data significantly extends the coverage of
the $M_{B}-z$ plane.
\label{fig2}}
 
\figcaption[fig3.eps]{The optical luminosities and redshifts of
radio-loud and -quiet quasars from the samples used in this paper,
showing the distribution of radio properties.  Empty circles denote
radio-quiet quasars, filled circles denote radio-louds.
\label{fig3}}
 
\figcaption[fig4.eps]{The radio-luminosity - optical luminosity plane
for quasars with $0.3 \le z \le 1.5$ used in this paper. Note that
there is no evidence for a correlation between radio and optical
luminosity for the quasars which have been detected in the radio, and
also that the radio luminosity distribution appears to be
bimodal.
\label{fig4}}
 
\figcaption[fig5.eps]{The radio-luminosity - optical luminosity plane
for quasars with $1.5 \le z \le 2.5$ used in this paper.
\label{fig5}}
 
\figcaption[fig6.eps]{Fraction of radio-loud quasars as a function of
redshift using data from this paper, together with the 2 parameter
evolution model (solid line indicates best-fit, dashed lines indicate
1$\sigma$ errors), and the no-evolution model (horizontal dot-dashed
line).  Also plotted is the fraction as determined from the PG sample
in Kellerman et al. (1989).
\label{fig6}}
 
\figcaption[fig7.eps]{Fraction of radio-loud quasars as a function of
absolute magnitude in the redshift ranges $0.3 \le z \le 1.3$ and $1.3
< z \le 2.5$ using data from this paper, the LBQS and MPM. The filled
circles show the fraction of radio-louds in the lower redshift bin,
and the empty circles show this fraction in the higher redshift bin.
The error bars show the 1 $\sigma$ errors. Also shown is the model fit
to the probability of a quasar being radio-loud as discussed in
Section 6.
\label{fig7}}
 
\figcaption[fig8.eps]{Results of Monte-Carlo simulation on the
probability of a quasar being radio-loud as a function of absolute
magnitude at $z=1$. Note that although we assumed that the intrinsic
disitribution is constant with respect to $M_{B}$ (horizontal line),
the observed distribution rises with increasing luminosity.
\label{fig8}}
 
\figcaption[fig9.eps]{The optical luminosity functions for radio-quiet
quasars (dashed lines) and radio-loud quasars (solid lines), estimated
from the derived likelihood of a quasar being radio-loud in Section
5.3. The lines are plotted for 4 different redshifts, from left to
right: $z=0.5, 1, 1.5, 2$.
\label{fig9}}

\clearpage
%table 1
%apj version of a table...

\begin{deluxetable}{lrlllrrr}
\footnotesize 
\tablecaption{5-GHz radio fluxes and positions for the
Edinburgh quasars. Where no radio source was detected within 5$''$ of
the quasar's optical position we quote the optical position and the
3-sigma noise measured at the centre of the map as an upper limit to
the flux density. Radio positions are accurate to 0.1$''$.
Colours, $B$ magnitudes and redshifts are taken from Goldschmidt
(1993). The 8.4 GHz flux from Hooper et al. (1995) is given in column
8 if the object has been included in that study. The symbol \dag
~indicates that the quasar has already been detected by Parkes, MIT,
Green Bank, Texas or Molonglo.
\label{tbl-1}}
\tablewidth{0pt}
\tablehead{
\colhead{Object} & 
\colhead{RA (B1950)} & 
\colhead{Dec (B1950)} & 
\colhead{B} & 
\colhead{U-B} & 
\colhead{z} & 
\colhead{5 GHz flux (mJy)} & 
\colhead{$8.4$ GHz flux (mJy)}}
% &   \colhead{RA (B1950)} 
% & \colhead{Dec (B1950)} & & & & \colhead{(Jy)}  \
\startdata
EQS B1228$-$0412  & 12 28 28.5   & $-$04 12 02  & 18.1 &  $-$0.5 &  0.658 & $<$0.3 &\nl
EQS B1228$-$0130 & 12 28 17.1    & $-$01 30 31  & 18.0 &  $-$1.2 &  0.706 & $<$0.3 & \nl
EQS B1229$-$0207$^{\dag}$  & 12 29 25.9   & $-$02 07 32.0  & 17.8 &  $-$0.9 & 1.045 & 966.6 & 284.0 \nl
EQS B1230$-$0015$^{\dag}$  & 12 30 30.24 & $-$00 15 01.7 & 17.7 &  $-$1.0 & 0.470 & 83.0 & 63.0 \nl 
EQS B1235$+$0216  & 12 35 39.7   & $+$02 16 48  & 18.2 &  $-$0.4 &  0.672 & $<$0.235 & $<$0.2 \nl 
\nl
EQS B1236$+$0128  & 12 36 38.0   & $+$01 28 42  & 18.0 &  $-$1.2 &  1.258 & $<$0.3 & $<$0.4 \nl 
EQS B1237$-$0435  & 12 37 41.7   & $-$04 35 03  & 18.4 &  $-$0.7 &  0.810 & $<$0.3 &  \nl 
EQS B1237$-$0359  & 12 37 05.7   & $-$03 59 22  & 17.9 &  $-$0.6 &  0.371 & $<$0.3 &  \nl 
EQS B1237$+$0204  & 12 37 58.4   & $+$02 04 44  & 17.7 &  $-$0.5 &  0.665 & $<$0.2 & $<$0.4 \nl 
EQS B1240$+$0224  & 12 40 13.9 & $+$02 24 43  & 17.9 &  $-$0.7 &  0.790 & 14.1 & 7.2 \nl 
\nl
EQS B1242$-$0123  & 12 42 22.1   & $-$01 23 10  & 18.1 &  $-$0.7 &  0.489 & $<$0.1 & $<$0.28 \nl 
EQS B1243$-$0026  & 12 43 39.4 & $-$00 26 10 & 17.1 &  $-$1.0 &  0.650 & 0.81 &  \nl 
EQS B1245$-$0333  & 12 45 00.4   & $-$03 33 47  & 16.5 &  $-$0.9 &  0.379 & $<$0.2 &  \nl 
EQS B1246$-$0430  & 12 46 17.0   & $-$04 30 03  & 17.8 &  $-$0.9 &  0.531 & $<$0.3 &  \nl 
EQS B1247$-$0213  & 12 47 13.2   & $-$02 13 09  & 18.5 &  $-$1.0 &  1.312 & $<$0.2 & $<$0.3 \nl 
\nl
EQS B1252$+$0200  & 12 52 46.4 & $+$02 00 29  & 15.4 &  $-$1.4 &  0.345 & 0.80 &  \nl 
EQS B1253$-$0002  & 12 53 16.4   & $-$00 02 17  & 17.7 &  $-$0.6 &  0.721 & $<$0.2 &  \nl 
EQS B1254$+$0206  & 12 54 33.3   & $+$02 06 52  & 17.3 &  $-$0.6 &  0.421 & $<$0.3 &  \nl 
EQS B1255$-$0143  & 12 55 40.9   & $-$01 43 08  & 17.7 &  $-$0.5 &  0.410 & $<$0.2 &  \nl 
EQS B1257$-$0140  & 12 57 18.2   & $-$01 40 58  & 17.7 &  $-$0.6 &  0.448 & $<$0.2 &  \nl 
\nl
%\nl
%\nl
%\nl
%\nl
EQS B1303$+$0205$^{\dag}$  & 13 03 21.05 & $+$02 05 32.3 & 17.1 &  $-$0.6 &  0.74 & 47.8 &\nl 
            & 13 03 20.36 & $+$02 05 11.0 &       &          &        & 40.07 &  \nl
EQS B1305$+$0100  & 13 05 20.8   & $+$01 00 22  & 17.8 &  $-$0.3 &  0.763 & $<$0.2 &  \nl 
EQS B1305$+$0230  & 13 05 42.5   & $+$02 30 10  & 17.2 &  $-$0.4 &  0.773 & $<$0.2 &  \nl  
EQS B1306$-$0213  & 13 06 32.4   & $-$02 13 18  & 17.6 &  $-$0.5 &  0.835 & $<$0.3 &  \nl 
\nl
EQS B1308$+$0109$^{\dag}$  & 13 08 47.78 & $+$01 09 15.2 & 17.9 &  $-$0.3 &  1.074 & 106.3 & 30.3 \nl  
EQS B1311$+$0217  & 13 11 53.6   & $+$02 17 07  & 17.1 &  $-$1.0 &  0.306& $<$0.3 & $<$0.5 \nl 
EQS B1313$-$0138  & 13 13 35.4   & $-$01 38 15  & 17.9 &  $-$0.7 &  0.406& $<$0.3 & $<$0.4 \nl 
EQS B1315$-$0410  & 13 15 14.0   & $-$04 10 14  & 17.7 &  $-$0.7 &  0.469& $<$0.2 &  \nl 
EQS B1315$+$0002  & 13 15 11.1   & $+$00 02 56  & 18.1 &  $-$0.7 &  0.917& $<$0.2 &  \nl 
\nl
EQS B1315$+$0140  & 13 15 41.9   & $+$01 40 37  & 18.1 &  $-$0.7 &  0.689& $<$0.2 & 0.34  \nl 
EQS B1316$-$0734  & 13 16 48.4   & $-$07 34 43  & 16.8 &  $-$0.7 &  0.538& $<$0.2 &  \nl 
EQS B1316$+$0023  & 13 16 06.5   & $+$00 23 21  & 18.0 &  $-$0.7 &  0.490& $<$0.2 & $<$0.3 \nl 
EQS B1317$-$0033$^{\dag}$  & 13 17 04.70 & $-$00 33 56.5 & 18.4 &  $-$0.8 &  890.0 &  678.7 &  \nl 
EQS B1319$+$0033  & 13 19 32.6   & $+$00 33 40  & 18.1 &  $-$0.7 &  0.535& $<$0.2 & $<$0.4 \nl 
\nl
\nl
\nl
\nl
EQS B1320$-$0006  & 13 20 49.9   & $-$00 06 17  & 18.4 &  $-$0.5 &  1.388& $<$0.2 & $<$0.4 \nl 
EQS B1321$-$0549  & 13 21 38.68 & $-$05 49 01.2 & 16.8 &  $-$0.6 &  0.732&  29.3 &  \nl 
EQS B1324$+$0126  & 13 24 50.9   & $+$01 26 55  & 18.3 &  $-$0.8 &  0.864& $<$0.2 &  \nl 
EQS B1328$-$0231  & 13 28 38.5   & $-$02 31 46  & 18.7 &  $-$0.7 &  1.240& $<$0.2 &  \nl 
EQS B1328$+$0205  & 13 28 58.7   & $+$02 05 12  & 18.2 &  $-$0.4 &  0.692& $<$0.2 & $<$0.4\nl 
\nl
%\enddata
%\end{deluxetable}
%\setcounter{table}{0}
%\begin{deluxetable}{lrlllrrr}
%\footnotesize
%\tablecaption{(cont.)}
%%\label{tbl-1}}
%\tablewidth{0pt}
%\tablehead{
%\colhead{Object} & 
%\colhead{Position} & 
%\colhead{B} & 
%\colhead{U-B} & 
%\colhead{z} & 
%\colhead{5 GHz flux} & 
%\colhead{$8.4$ GHz flux}}
% &   \colhead{RA (B1950)} 
% & \colhead{Dec (B1950)} & & & & \colhead{(Jy)}  \
%\startdata
EQS B1329$-$0615  & 13 29 09.4   & $-$06 15 21  & 17.6 &  $-$0.46 &  0.718& $<$0.2  & \nl 
EQS B1329$-$0007  & 13 29 57.1   & $-$00 07 55  & 18.3 &  $-$0.68 &  0.962& $<$0.2 &  \nl 
EQS B1334$-$0232  & 13 34 37.9   & $-$02 32 37  & 17.7 &  $-$0.38 &  0.732& $<$0.3  & $<$0.4\nl
EQS B1334$+$0053  & 13 34 49.2   & $+$00 53 27  & 18.3 &  $-$0.45 &  0.647& $<$0.2  & $<$1.8\nl 
EQS B1335$-$0611$^{\dag}$  & 13 35 31.25 & $-$06 11 56.8 & 17.8 &  $-$0.42 &  0.620&   1021.8  & \nl 
\nl
EQS B1335$-$0241  & 13 35 01.6   & $-$02 41 54  & 18.0 &  $-$0.3 &  0.610& $<$0.2  & $<$0.4\nl 
EQS B1335$+$0222$^{\dag}$  & 13 35 06.93 & $+$02 22 12.5  & 18.0 &  $-$0.6 &  1.354& 107.6  & \nl 
EQS B1337$-$0146  & 13 37 17.0   & $-$01 46 08  & 17.9 &  $-$0.5 &  1.010& $<$0.2  & $<$0.2\nl 
EQS B1338$-$0030  & 13 38 10.4   & $-$00 30 07  & 17.1 &  $-$0.6 &  0.389& $<$0.2  & $<$0.3 \nl 
EQS B1339$-$0648  & 13 39 49.7   & $-$06 48 05  & 18.3 &  $-$0.9 &  1.220& $<$0.2  & \nl 
\nl
EQS B1339$-$0526  & 13 39 41.4   & $-$05 26 12  & 18.1 &  $-$0.5 &  1.252& $<$0.3  & \nl 
EQS B1339$-$0459  & 13 39 39.3   & $-$04 59 49  & 18.2 &  $-$0.8 &  0.884& $<$0.7  & \nl 
EQS B1340$-$0020  & 13 40 12.2   & $-$00 20 38  & 18.2 &  $-$0.6 &  0.792& $<$0.2  & $<$0.4\nl 
EQS B1340$+$0107  & 13 40 25.8   & $+$01 07 03  & 17.9 &  $-$0.7 &  1.067& $<$0.2  & $<$0.3 \nl 
EQS B1341$-$0357  & 13 41 34.4   & $-$03 57 47  & 17.3 &  $-$0.8 &  0.835& $<$0.2  & \nl 
\nl
EQS B1343$-$0607  & 13 43 23.7   & $-$06 07 44  & 17.7 &  $-$0.7 &  1.012& $<$0.2  & \nl 
EQS B1343$-$0221  & 13 43 13.1   & $-$02 21 55  & 18.2 &  $-$0.3 &  0.509& $<$0.2  & 0.7 \nl 
EQS B1344$-$0227  & 13 44 38.1   & $-$02 27 36  & 18.2 &  $-$0.4 &  0.511& $<$0.2  & $<$0.4\nl 
EQS B1345$-$0000  & 13 45 17.8   & $-$00 00 23  & 18.0 &  $-$0.4 &  0.552& $<$0.2  & $<$0.5\nl 
EQS B1346$+$0007  & 13 46 48.3   & $+$00 07 55  & 18.3 &  $-$0.2 &  1.127& $<$0.2  & $<$0.2\nl 
\nl
EQS B1347$-$0026  & 13 47 00.2   & $-$00 26 11  & 17.6 &  $-$0.4 &  0.515& $<$0.2  & $<$0.3\nl 
EQS B1348$+$0118  & 13 48 55.1   & $+$01 18 28  & 17.7 &  $-$0.7 &  1.094& $<$0.2  & $<$0.3\nl 
EQS B1349$+$0057  & 13 49 59.2   & $+$00 57 39  & 18.0 &  $-$0.8 &  1.151& $<$0.3  & $<$0.3\nl 
EQS B1352$-$0043  & 13 52 51.4   & $-$00 43 00  & 18.4 &  $-$0.8 &  0.900& $<$0.3  & \nl 
EQS B1354$-$0233  & 13 54 53.8   & $-$02 33 03  & 17.1 &  $-$0.6 &  0.559& $<$0.6  & \nl 
\nl
EQS B1354$+$0117  & 13 54 17.2   & $+$01 17 22  & 17.9 &  $-$0.8 &  1.210& $<$1.5  & \nl 
EQS B1357$-$0227  & 13 57 31.4   & $-$02 27 02  & 17.6 &  $-$0.7 &  0.418& $<$0.3 &  \nl 
EQS B1358$+$0058  & 13 58 31.6   & $+$00 58 01  & 17.5 &  $-$0.6 &  0.664&  1.0 &  \nl 
EQS B1403$-$0304  & 14 03 35.2   & $-$03 04 56  & 17.2 &  $-$0.8 &  0.860& $<$0.2 &  \nl 
EQS B1406$-$0143  & 14 06 54.15  & $-$01 43 08.8 & 17.7 &  $-$0.7 &  0.644&  72.4 &  \nl 
\nl
EQS B1407$-$0722  & 14 07 52.80  & $-$07 22 32  & 18.1 &  $-$0.9 &  0.900& $<$0.2 &  \nl 
EQS B1407$-$0231  & 14 07 20.92 & $-$02 31 56.7 & 17.9 &  $-$0.7 &  1.263&  43.9 &  \nl 
                  & 14 07 24.20 & $-$02 31 24.4 &  &  &  &  29.0 & \nl
EQS B1411$-$0333  & 14 11 41.4   & $-$03 33 47  & 17.9 &  $-0.5$  &  0.860& $<$0.2 &  \nl 
EQS B1413$+$0107  & 14 13 16.6   & $+$01 07 51  & 17.4 &  $-$0.6 &  1.042& $<$0.2 &  \nl 
\nl
EQS B1414$-$0544  & 14 14 30.8   & $-$05 44 46  & 17.5 &  $-$0.6 &  0.419& $<$0.3 &  \nl 
EQS B1416$-$0518  & 14 16 33.6   & $-$05 18 59  & 18.5 &  $-$0.7 &  0.96 & 13.1 &  \nl 
EQS B1420$-$0053  & 14 20 05.9   & $-$00 53 09  & 17.7 &  $-$0.5 &  0.717& $<$0.3 &  \nl 
EQS B1421$-$0407  & 14 21 33.4   & $-$04 07 42 & 18.1 &  $-$0.4 &  0.650& $<$0.2 &  \nl 
EQS B1421$+$0108  & 14 21 57.3   & $+$01 08 32  & 18.1 &  $-$0.7 &  1.060& $<$0.3 &  \nl 
\nl
EQS B1423$-$0013  & 14 23 26.2   & $-$00 13 31 & 17.9 &  $-$0.6 &  1.078& $<$0.3 &  \nl 
EQS B1424$-$0007  & 14 24 24.6   & $-$00 07 30  & 16.5 &  $-$0.5 &  0.632& $<$0.3 &  \nl 
EQS B1429$-$0100  & 14 29 07.3   & $-$01 00 17 & 17.4 &  $-$0.7 &  0.661& $<$0.2 & $<$0.4\nl
EQS B1429$-$0036  & 14 29 09.4   & $-$00 36 58  & 18.5 &  $-$0.7 &  1.179& $<$0.4 & $<$0.8  \nl
EQS B1430$-$0046$^{\dag}$  & 14 30 10.0  & $-$00 46 04  & 17.7 &  $-$0.8 &  1.0 & 22.9 & 15.0 \nl 
            & 14 30 11.1 & $-$00 46 18   &       &          &       & 14.0 &  \nl
            & 14 30 09.1  & $-$00 45 53   &       &          &       & 20.6 &  \nl 
\nl
EQS B1437$-$0143  & 14 37 46.8   & $-$01 43 37  & 18.3 &  $-$0.6 &  0.718& $<$0.2 & $<$0.5 \nl 
EQS B1440$-$0234  & 14 40 38.4   & $-$02 34 40  & 17.4 &  $-$0.7 &  0.675& $<$0.3 & $<$0.3\nl
EQS B1440$+$0149  & 14 40 18.0   & $+$01 49 38  & 18.2 &  $-$0.8 &  1.170& $<$0.2 & $<$0.5 \nl
EQS B1446$+$0218  & 14 46 05.7   & $+$02 18 54  & 18.1 &  $-$0.5 &  0.668& $<$0.2 & $<$0.4\nl
\enddata
\end{deluxetable}

\clearpage

%table 2
\begin{table*}
\begin{center}
\begin{tabular}{lrrrrr}
RA (1950)& Dec (1950)  & $B$ & $z$ & $S_{\nu}$ & Offset\\
\tableline
 00 51 38 &  $-$27 26 25  &  18.9 &  0.689 &  6.5 &  3.3 \\
 00 49 28 &  $-$28 12 36  &  20.1 &  1.145 &  5.9 &  1.7 \\
 22 03 26 &  $-$18 50 17  &  19.0 &  0.620 & 6398.9 &  2.1 \\
 22 37 43 &  $-$39 18 35  &  19.1 &  0.713 &  2.4 &  4.7 \\
\end{tabular}
\end{center}
\tablenum{2}
\caption{Table of quasars from AAT survey (Boyle et al., 1990) which
have been detected in the NVSS (Condon et al., 1998). We present the
1950 coordinates of the optical position, the optical magnitudes, and
radio fluxes in mJy, and lastly the offset between the optical and
radio positions in arcseconds.
\label{tbl-2}}

\end{table*}

\clearpage

%table 3
\begin{table*}
\caption{ Table of quasars from the LBQS survey (Hewett et al., 1995)
which have been detected in FIRST (White et al., 1997). We present the
B1950 coordinates of the optical position, the optical magnitudes (in
B Johnson), and radio fluxes in mJy, and lastly the offset between the
optical and radio positions in arcsecs.
\label{tbl-3}}
\begin{center}
\begin{tabular}{lrrrrr}
RA (1950)& Dec (1950)  & $B$ & $z$ & $S_{\nu}$ & Offset\\
\tableline
 00 02 33 &  $-$01 49 28 &   18.8 & 1.709& 63.9 & 0.9 \\
 00 04 36 &  $+$00 36 46 &   17.9 & 0.316&  1.4 & 1.2 \\
 00 09 41 &  $-$01 48 10 &   17.8 & 1.072&  1.1 & 0.1 \\
 00 12 10 &  $-$00 16 59 &   18.3 & 1.524&  1.5 & 1.3 \\
 00 12 33 &  $-$00 24 42 &   18.7 & 1.701& 13.5 & 0.7 \\
 00 19 07 &  $+$00 22 03 &   18.7 & 0.313&  1.4 & 0.4 \\
 00 20 11 &  $-$02 02 29 &   18.4 & 0.691& 214.8 & 0.3 \\
 00 21 38 &  $-$01 00 25 &   18.2 & 0.764&  1.0 & 1.0 \\
 00 24 44 &  $+$00 20 47 &   18.0 & 1.228&  3.7 & 1.5 \\
 00 29 03 &  $-$01 52 56 &   18.7 & 2.382& 14.1 & 0.1 \\
 00 48 57 &  $+$00 25 32 &   18.2 & 1.187& 13.8 & 1.0 \\
 00 49 32 &  $+$00 19 21 &   16.5 & 0.397& 88.0 & 1.0 \\
 00 52 07 &  $-$00 15 04 &   17.8 & 0.647&  2.7 & 0.4 \\
 00 56 32 &  $-$00 09 19 &   17.8 & 0.717& 2416.0 &  0.3 \\
 00 59 32 &  $-$02 06 46 &   18.1 & 1.320&  1.8 & 0.5 \\
 01 07 40 &  $-$02 35 51 &   18.2 & 0.957& 188.5 &  3.0 \\
 02 49 22 &  $+$00 44 50 &   18.5 & 0.469&  8.1 & 0.7 \\
 02 51 07 &  $-$00 01 02 &   18.4 & 1.682&  7.6 & 1.1 \\
 02 56 06 &  $-$02 06 08 &   18.5 & 0.406&  1.1 & 1.4 \\
 02 56 32 &  $-$00 00 34 &   18.3 & 3.363&  2.6 & 1.4 \\
% 02 56 55 & $-$00 31 54 &   17.7 & 1.995&  9.6 & 8.9 \\
 02 56 55 &  $-$00 31 54 &   17.7 & 1.995& 225.0 &  0.5 \\
% 02 56 55 & $-$00 31 54 &   17.7 & 1.995& 12.7 & 9.2 \\
 02 57 03 &  $+$00 25 42 &   16.9 & 0.531&  0.8 & 0.1 \\
 22 31 26 &  $-$00 48 46 &   17.6 & 1.209&  1.0 & 1.5 \\
 22 35 01 &  $+$00 54 58 &   18.6 & 0.528&  1.1 & 1.0 \\
 22 45 05 &  $-$00 55 44 &   17.5 & 0.800&  1.6 & 1.8 \\
 23 51 35 &  $-$00 36 30 &   18.5 & 0.460& 346.1 &  0.6 \\
 23 59 16 &  $-$02 16 23 &   18.8 & 2.816& 25.0 & 6.3 \\
 23 59 49 &  $-$00 21 26 &   18.6 & 0.809&  3.9 & 0.9 \\
\end{tabular}
\end{center}
\tablenum{3}
\end{table*}

\begin{figure}
\vspace {21.5truecm}
\includegraphics{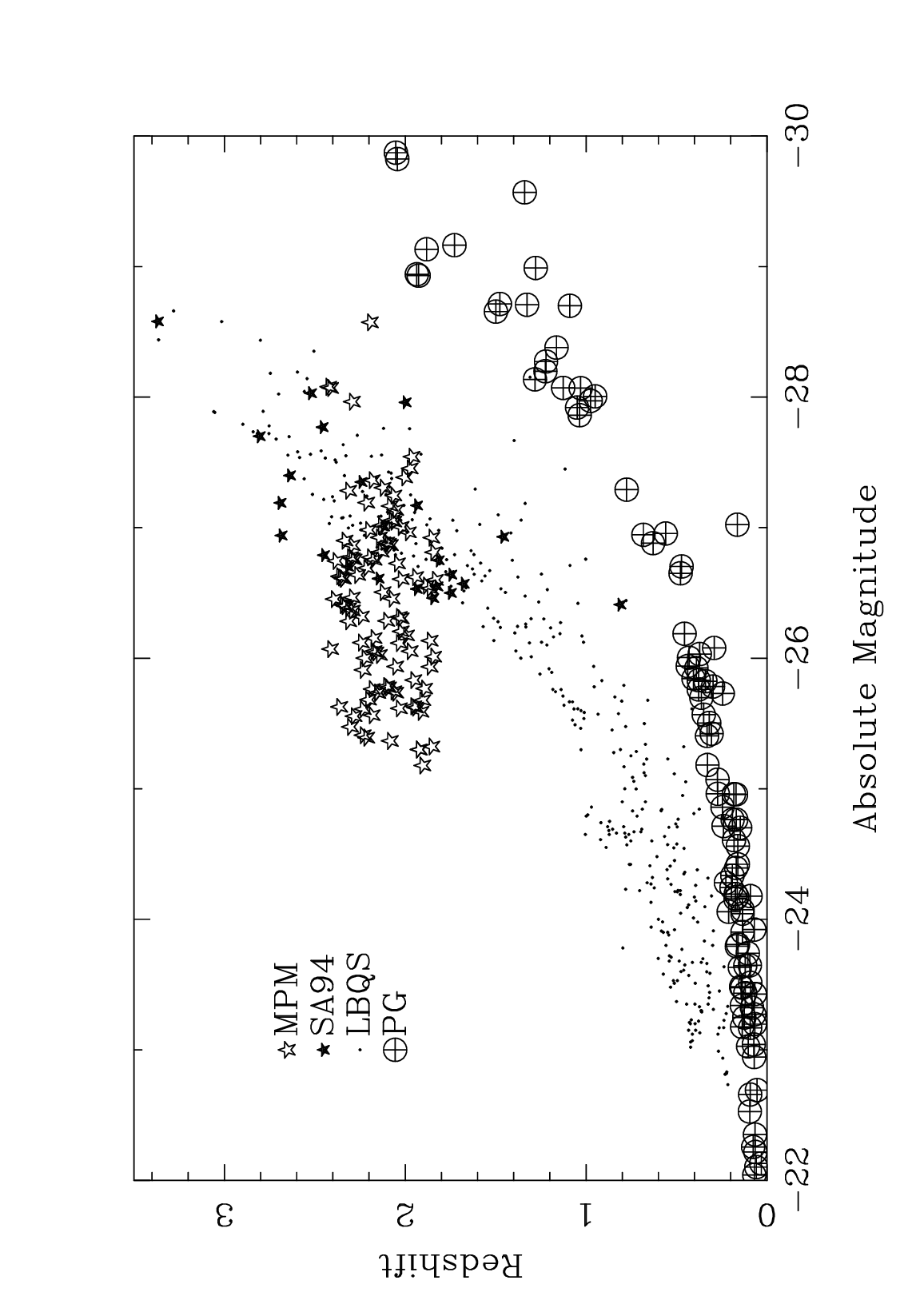}
\hspace{7.2truecm}
{\Large {Figure 1}}
\end{figure}

\begin{figure}
\vspace {21.5truecm}
\includegraphics{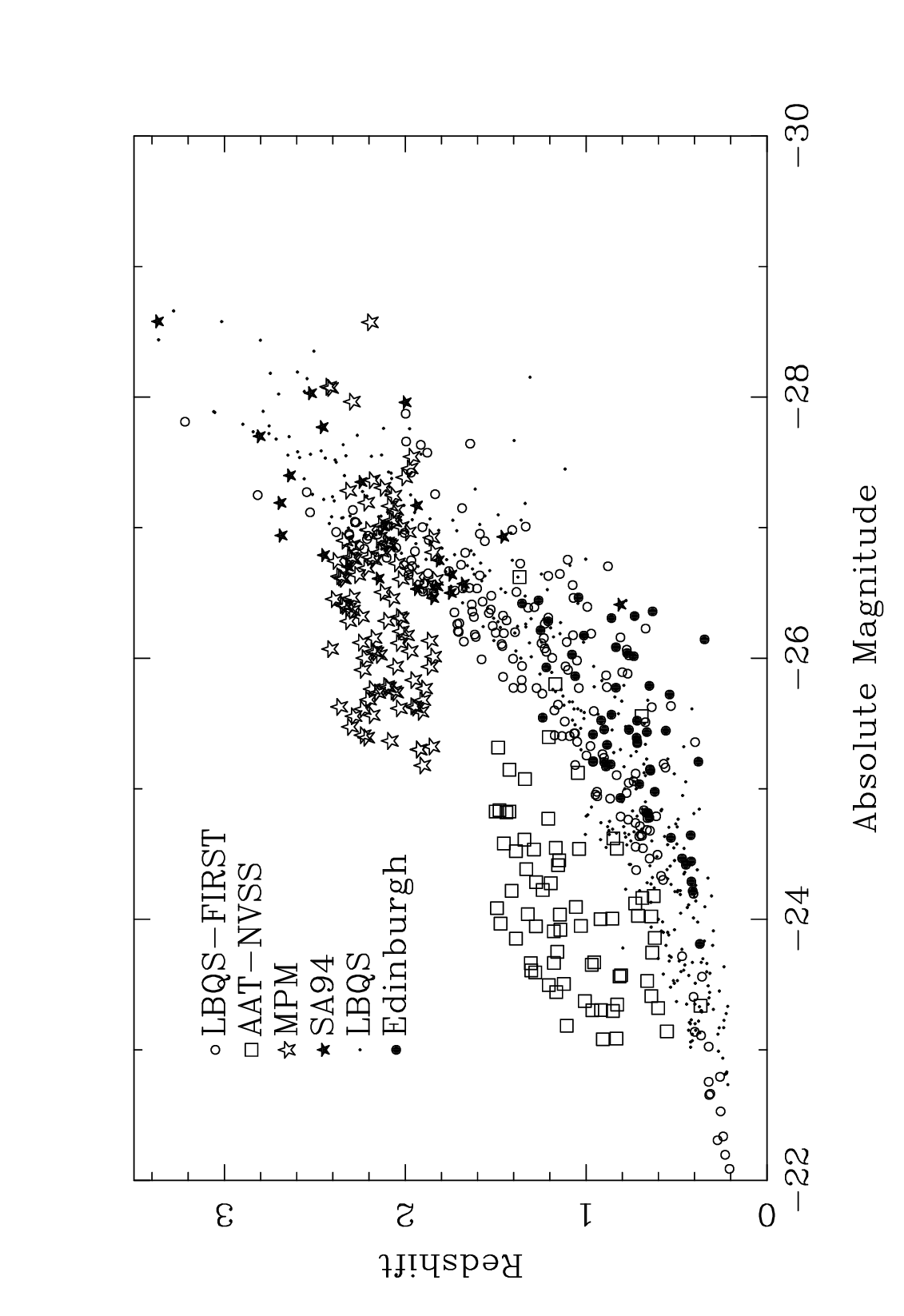}
\hspace{7.2truecm}
{\Large {Figure 2}}
\end{figure}

\begin{figure}
\vspace {21.5truecm}
\includegraphics{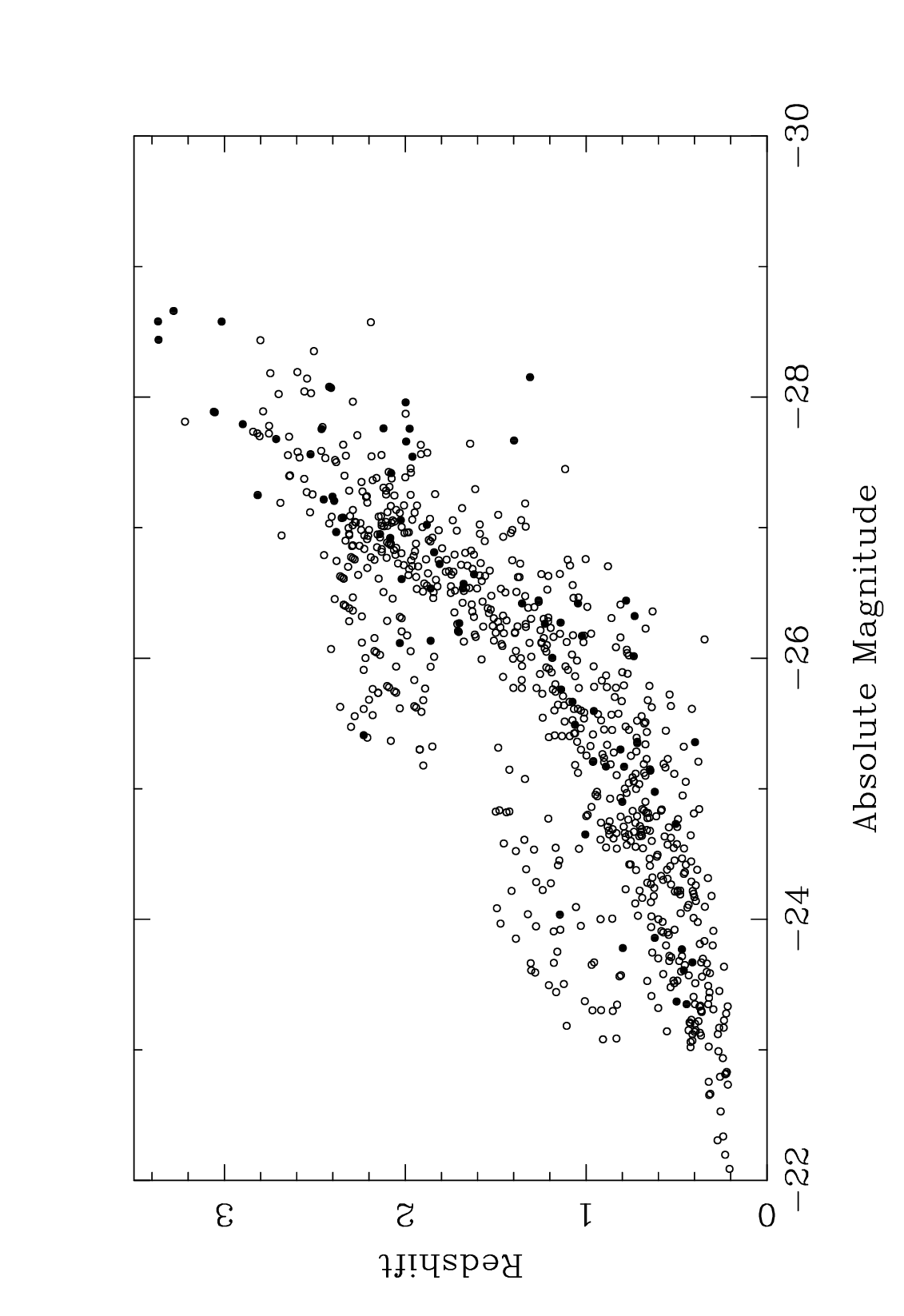}
\hspace{7.2truecm}
{\Large {Figure 3}}
\end{figure}

\begin{figure}
\vspace {21.5truecm}
\includegraphics{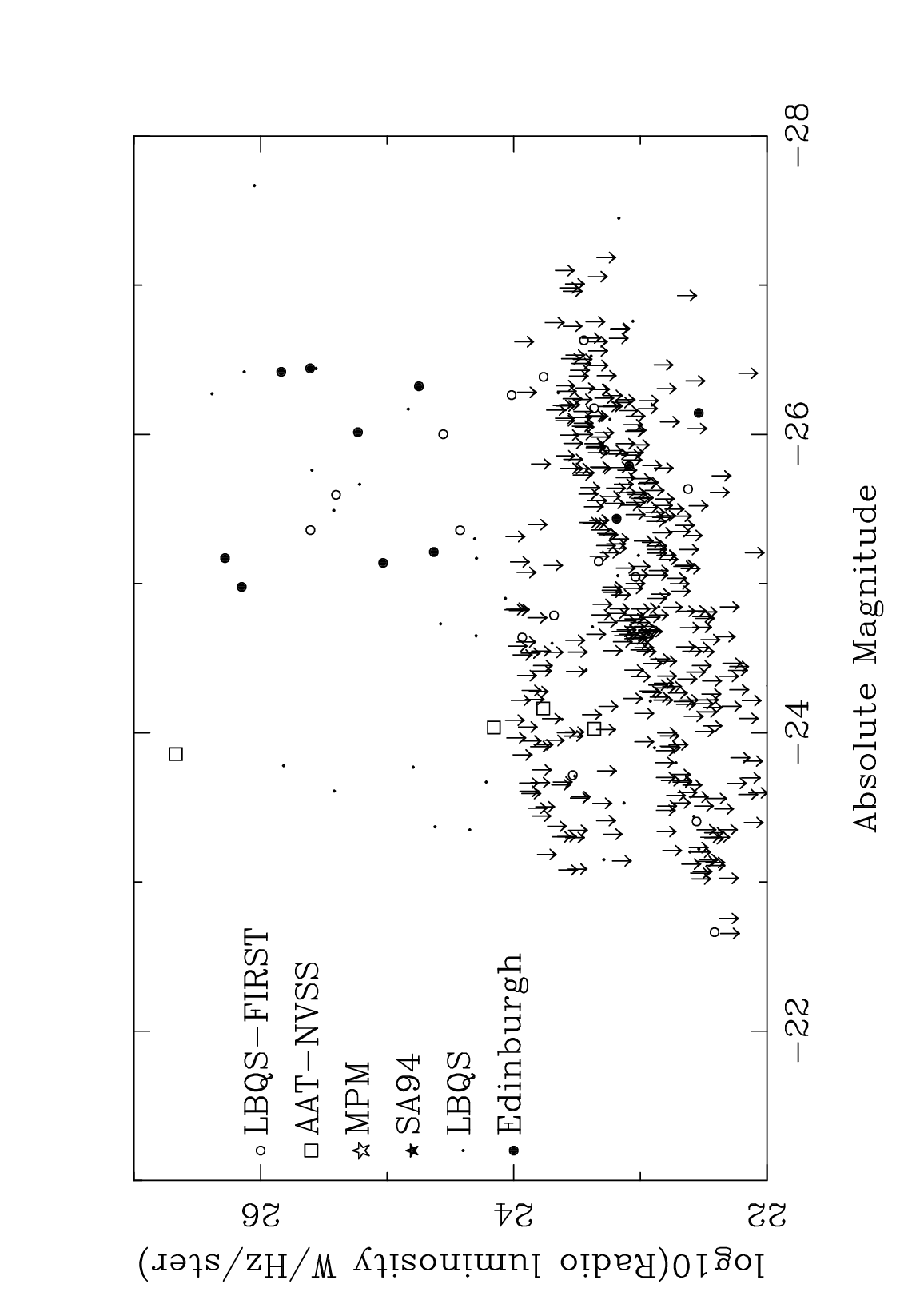}
\hspace{7.2truecm}
{\Large {Figure 4}}
\end{figure}

\begin{figure}
\vspace {21.5truecm}
\includegraphics{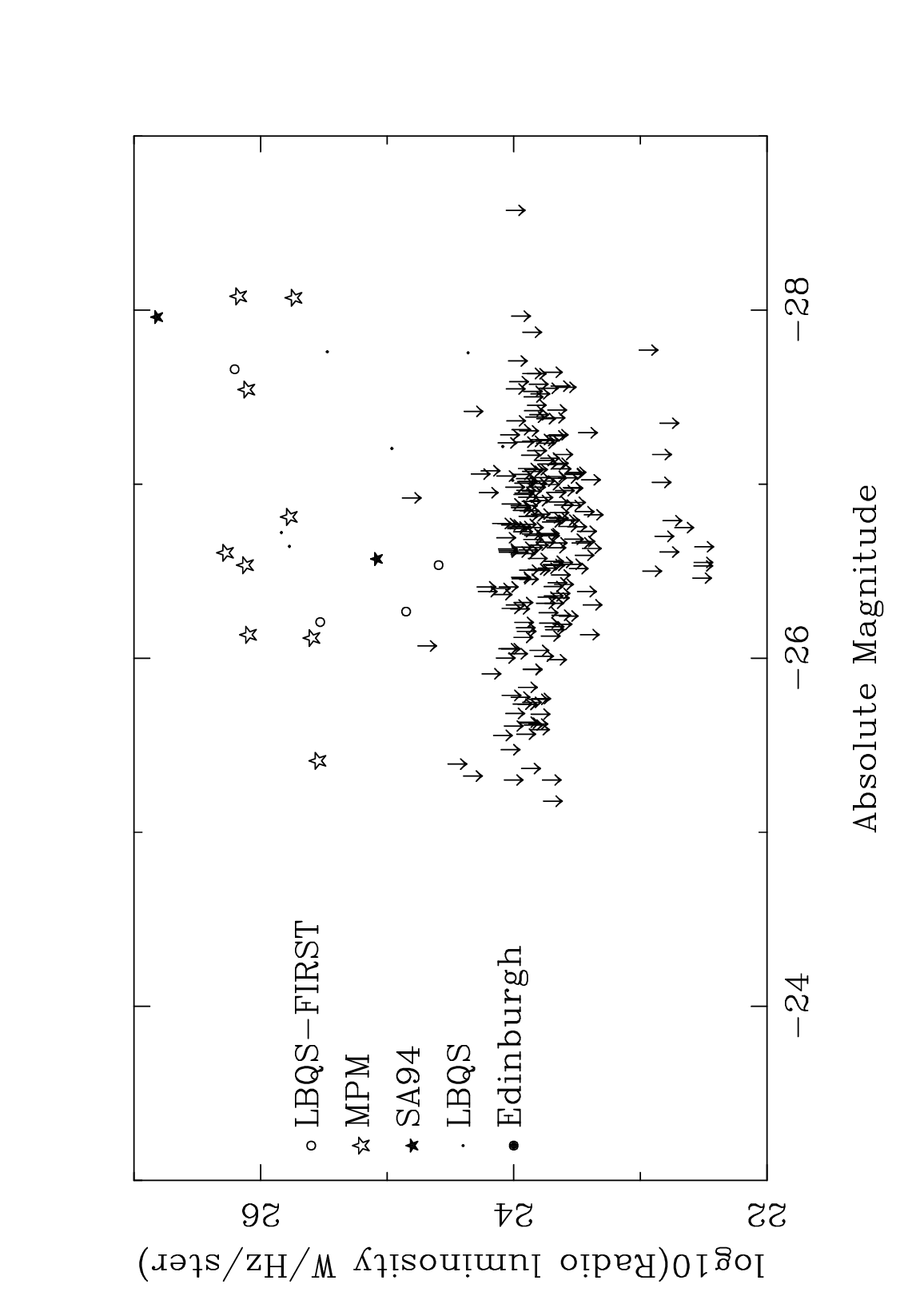}
\hspace{7.2truecm}
{\Large {Figure 5}}
\end{figure}

\begin{figure}
\vspace {21.5truecm}
\includegraphics{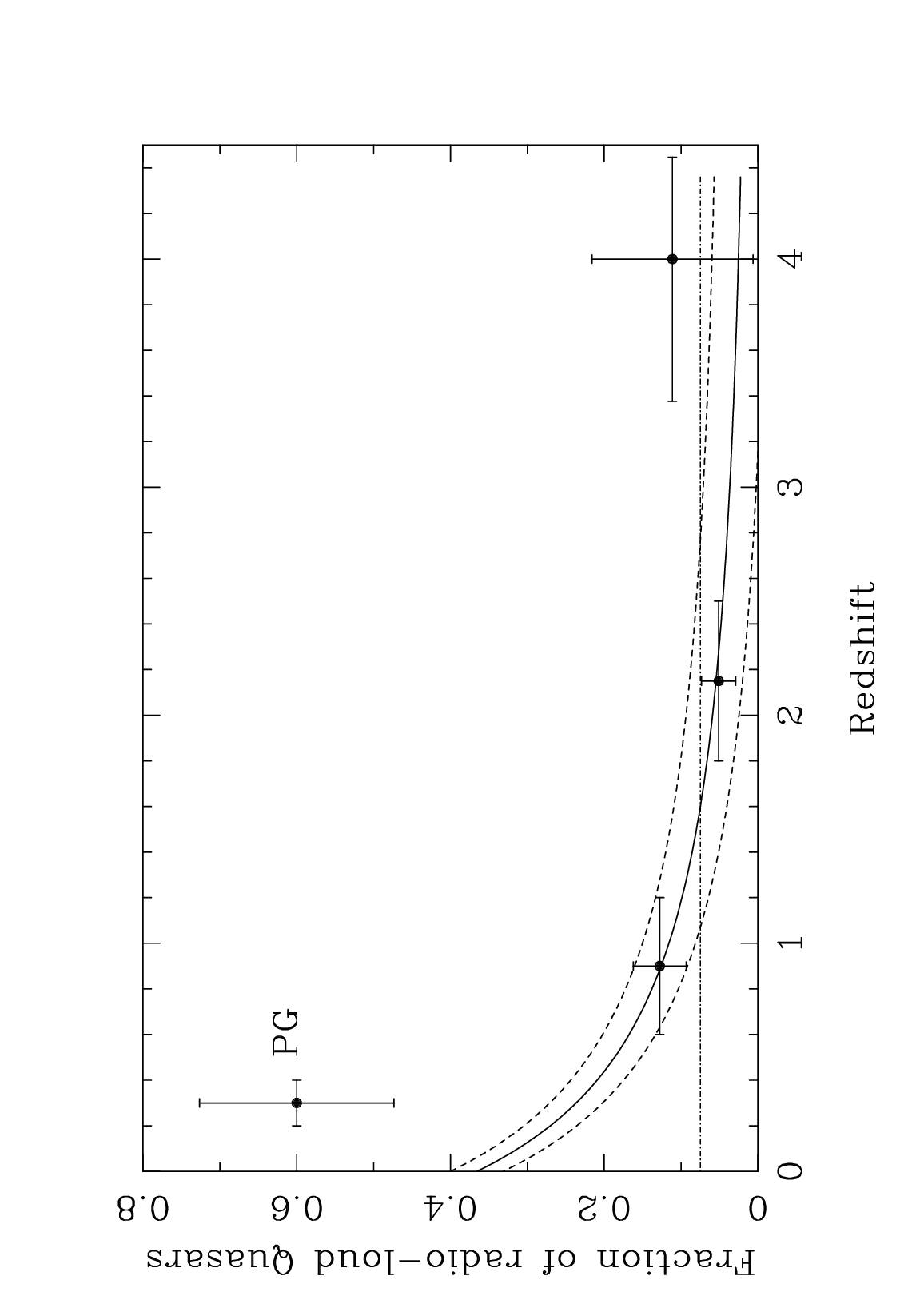}
\hspace{7.2truecm}
{\Large {Figure 6}}
\end{figure}

\begin{figure}
\vspace {21.5truecm}
\includegraphics{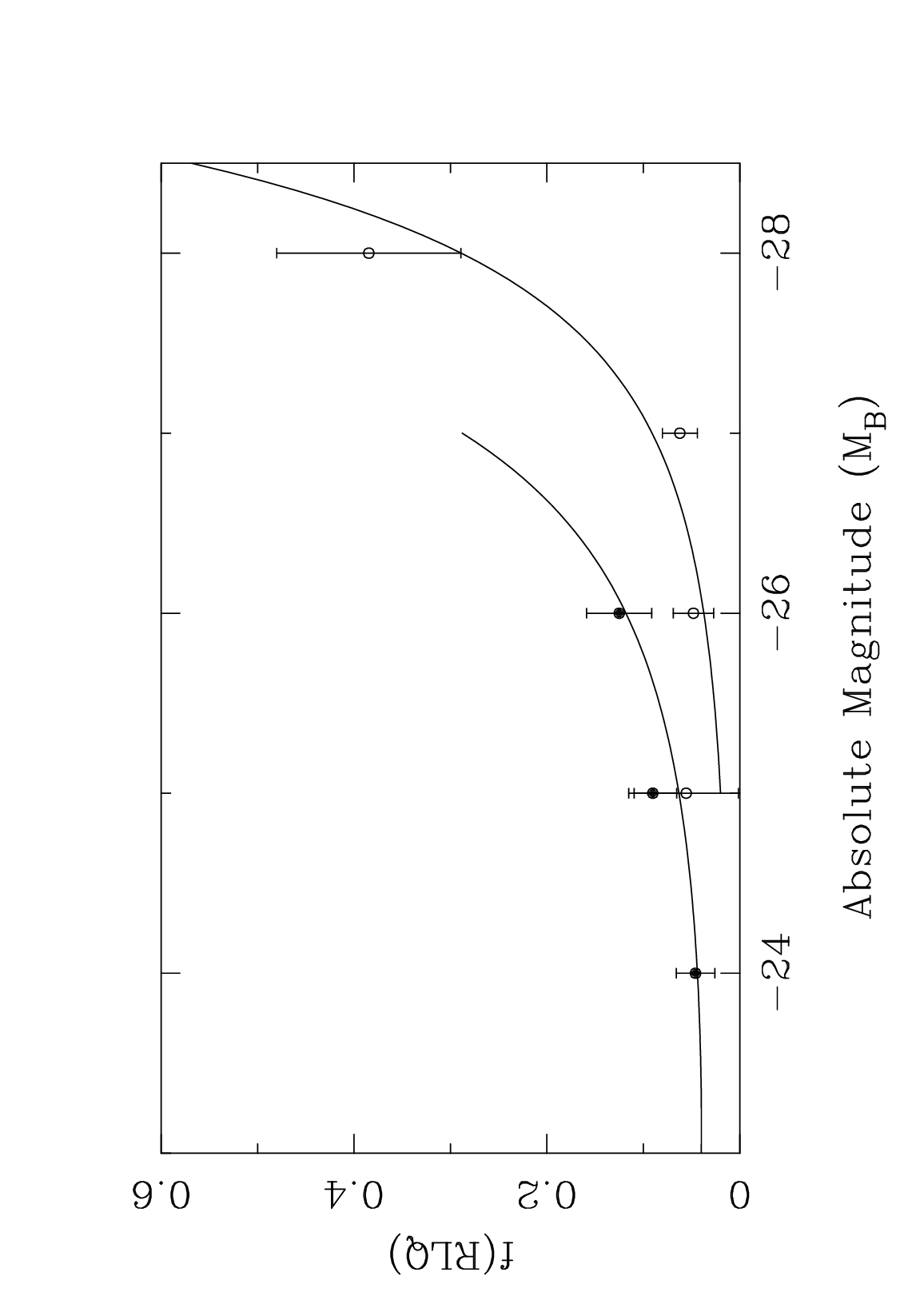}
\hspace{7.2truecm}
{\Large {Figure 7}}
\end{figure}

\begin{figure}
\vspace {21.5truecm}
\includegraphics{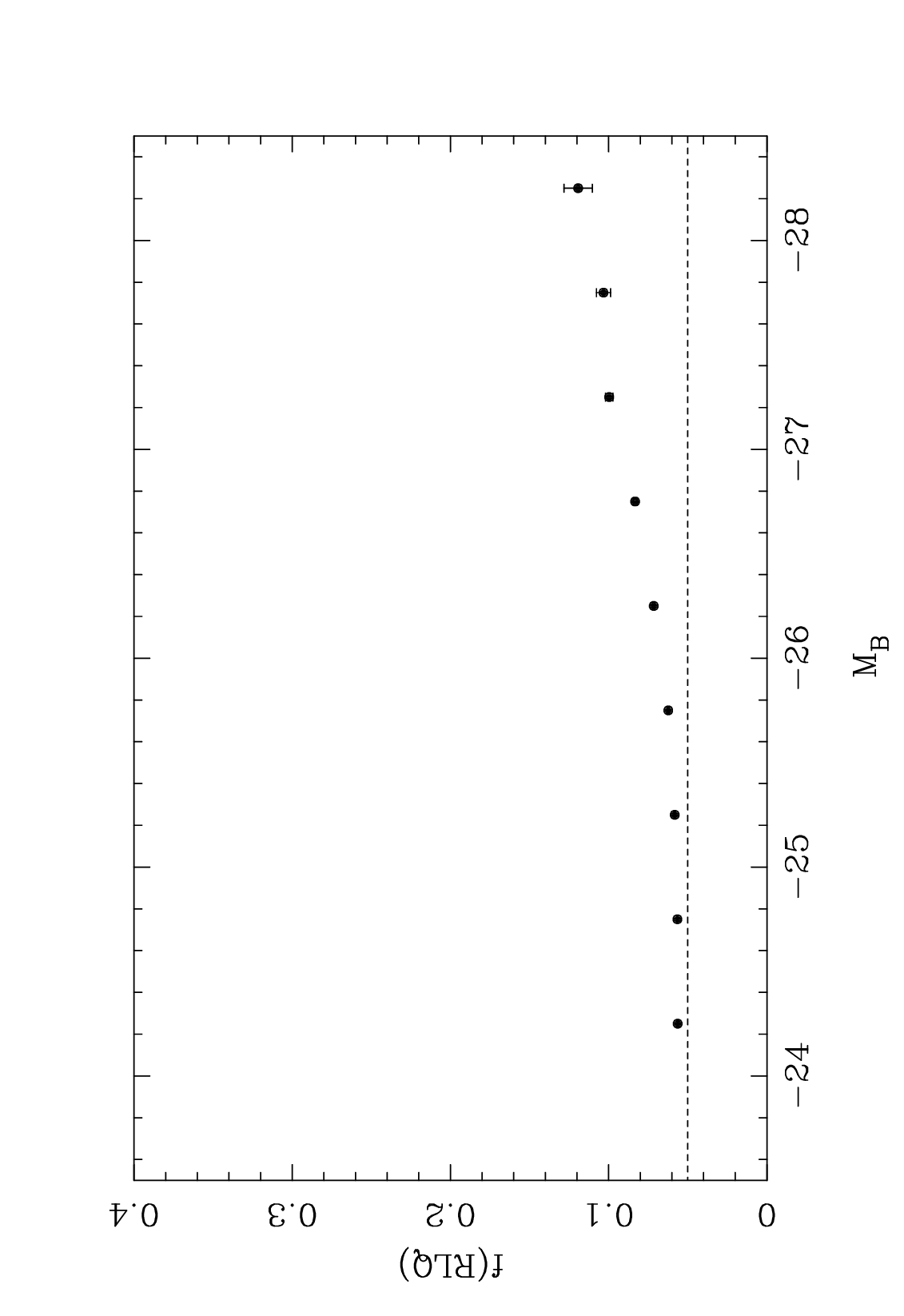}
\hspace{7.2truecm}
{\Large {Figure 8}}
\end{figure}

\begin{figure}
\vspace {21.5truecm}
\includegraphics{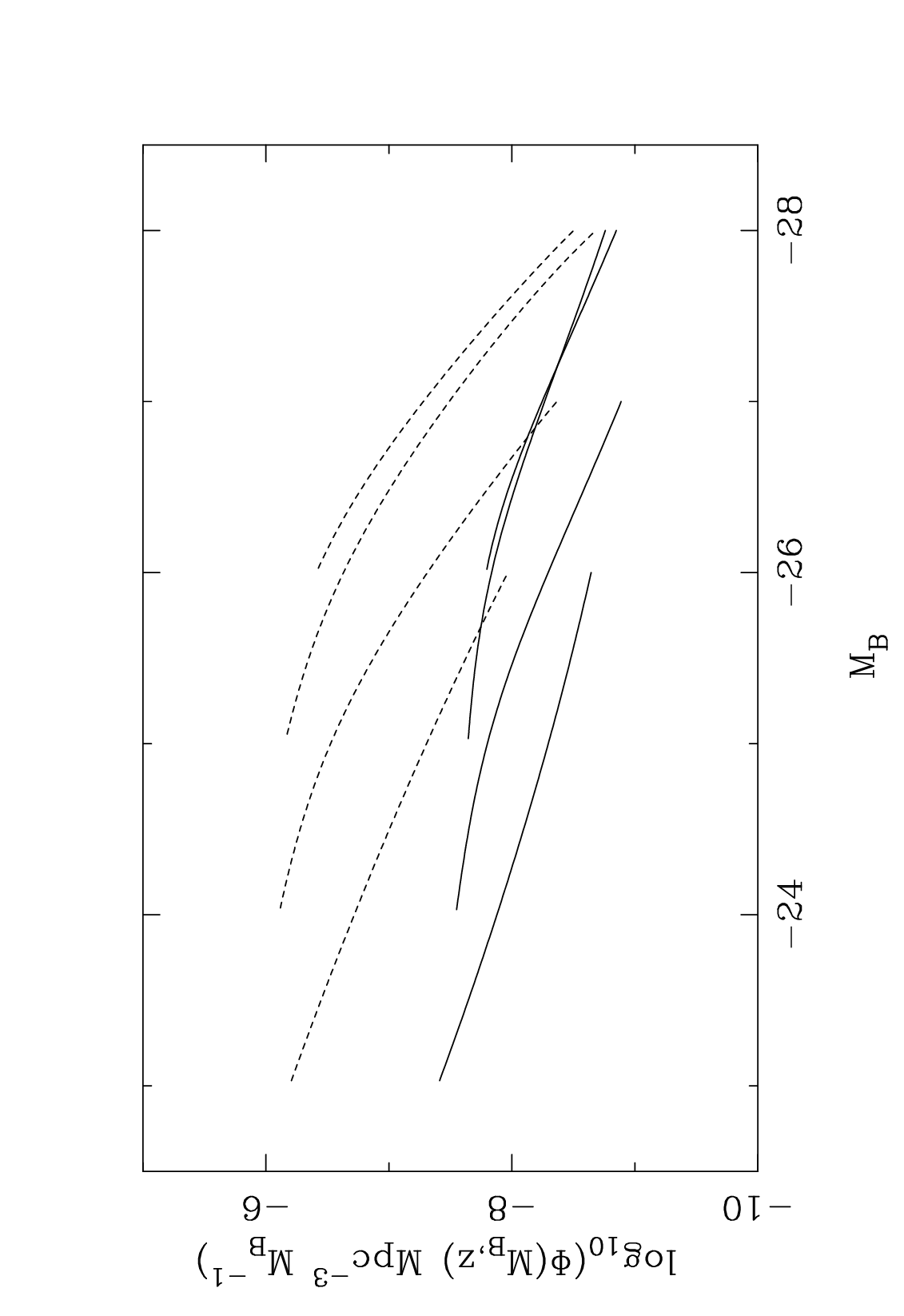}
\hspace{7.2truecm}
{\Large {Figure 9}}
\end{figure}

\end{document}